\documentclass[aps,prd,onecolumn,12pt,amssymb,showpacs,superscriptaddress,nofootinbib,longbibliography]{revtex4-1}

\usepackage{amsfonts}
\usepackage{latexsym}
\usepackage{yfonts}
\usepackage{amsmath}
\usepackage{amssymb}
\usepackage{amsthm}
\usepackage{mathrsfs}
\usepackage{upgreek}
\usepackage{bm}
\usepackage{dcolumn}
\usepackage{epsfig}
\usepackage{graphicx}
\usepackage{latexsym}
\usepackage{rotating}
\usepackage{color}
\usepackage[dvipsnames]{xcolor}
\usepackage{soul}
\usepackage[english]{babel}
\soulregister\cite7
\soulregister\ref7
\soulregister\pageref7

\usepackage{lmodern}
\usepackage[T1]{fontenc}
\usepackage[utf8]{inputenc}

\usepackage{times}
\usepackage{setspace}

\usepackage{color}
\usepackage[unicode=true,pdfusetitle,
 bookmarks=true,bookmarksnumbered=false,bookmarksopen=false,
 breaklinks=false,pdfborder={0 0 0},backref=false,colorlinks=true]
 {hyperref}
\hypersetup{linkcolor=[rgb]{0.1,0.4,0.1}, linktoc=page}
\hypersetup{citecolor=[rgb]{0.6,0,0.1}}
\hypersetup{urlcolor=[rgb]{0.1,0,0.6}}

\DeclareTextFontCommand{\emph}{\sl}

\usepackage{setspace}
\usepackage[hang]{footmisc}
\usepackage{lipsum}

\makeatletter
\renewcommand{\overset}[3][0ex]{%
  \mathrel{\mathop{#3}\limits^{
    \vbox to#1{\kern-2\ex@
    \hbox{$\scriptstyle#2$}\vss}}}}
\makeatother

\definecolor{capri}{rgb}{0.0, 0.75, 1.0}

\definecolor{ggreen}{rgb}{0.05,0.45,0.1}

\AtBeginDocument{%
    \newwrite\bibnotes
    \def\bibnotesext{Notes.bib}
    \immediate\openout\bibnotes=\jobname\bibnotesext
    \immediate\write\bibnotes{@CONTROL{REVTEX41Control}}
    \immediate\write\bibnotes{@CONTROL{%
    apsrev41Control,author="08",editor="1",pages="1",title="0",year="1"}}
     \if@filesw
     \immediate\write\@auxout{\string\citation{apsrev41Control}}%
    \fi
}%

\renewcommand{\emph}[1]{\textit{#1}}

\begin{document}

\title{Quasilocal conservation laws in cosmology: a first look}

\thanks{Essay awarded Honorable Mention in the Gravity Research Foundation 2020 Awards for Essays on Gravitation. Submitted for publication in the October 2020 Special Issue of the International Journal of Modern Physics D.}

\author{Marius Oltean}
\email[]{oltean@ice.cat}
\thanks{corresponding author.}

\affiliation{\mbox{Department of Operations, Innovation and Data Sciences,}\\\vspace{-0.075cm}
\mbox{ESADE Business School, Av. Torreblanca 59, 08172 Sant Cugat (Barcelona), Spain}}

\affiliation{\mbox{Institute of Space Studies of Catalonia (IEEC),}\\\vspace{-0.075cm}
\mbox{Carrer del Gran Capit\`{a}, 2-4, Edifici Nexus, despatx 201, 08034 Barcelona, Spain}}

\author{Hossein Bazrafshan Moghaddam}
\email[]{hbazrafshan@um.ac.ir}

\affiliation{\mbox{Department of Physics, Faculty of Science, Ferdowsi University of Mashhad, Mashhad, Iran}}

\author{Richard J. Epp}
\email[]{rjepp@uwaterloo.ca}

\affiliation{\mbox{Department of Physics and Astronomy, University of Waterloo,}\\\vspace{-0.075cm}
\mbox{200 University Avenue West, Waterloo, Ontario N2L 3G1, Canada}}

\date{\today}

\begin{abstract}

\begin{spacing}{1}

Quasilocal definitions of stress-energy-momentum---that is, in the form of boundary densities (in lieu of local volume densities)---have proven generally very useful in formulating and applying conservation laws in general relativity. In this essay, we take a first basic look into applying these to cosmology, specifically using the Brown-York quasilocal stress-energy-momentum tensor for matter and gravity combined. We compute this tensor and present some simple results for a flat FLRW spacetime with a perfect fluid matter source.
\end{spacing}

\end{abstract}

\maketitle

\pagebreak

In cosmology, as in many other situations in general relativity requiring
computational convenience, the notion of gravitational energy-momentum
is often treated effectively, in analogy with matter (non-gravitational)
energy-momentum, as a local concept. For example, in the Einstein
equation\footnote{\setstretch{0.9}We work in the $(-+++)$ signature of spacetime,
in geometrized units ($G=1=c$), and follow the conventions of Wald
\cite{wald_general_1984}. In particular, Latin letters are used for
abstract spacetime indices ($a,b,c...=0,1,2,3$). Furthermore we denote
by $\bm{\epsilon}_{\mathscr{U}}$ the volume form of any manifold
$\mathscr{U}$.}, 
\begin{equation}
G_{ab}+\Lambda g_{ab}=8\pi T_{ab}\,,\label{eq:EE}
\end{equation}
the typical interpretation given to the cosmological constant term
follows by moving it to the RHS: $G_{ab}=8\pi(T_{ab}+T_{ab}^{(\Lambda)})$,
with $T_{ab}^{(\Lambda)}=-\frac{1}{8\pi}\Lambda g_{ab}$ interpreted
as playing the role of an effective local stress-energy-momentum of
the ``gravitational vacuum'', with a constant local energy volume
density $\rho_{\Lambda}=T_{00}^{(\Lambda)}$ and equal but negative
local pressure $p_{\Lambda}=-\rho_{\Lambda}$.

Yet it has long been understood that, generally, gravitational energy-momentum
cannot be treated fundamentally as a local concept in general relativity. There
are many reasons for this, but the simplest explanation comes directly
from the equivalence principle (see e.g. section 20.4 of \cite{misner_gravitation_1973}):
if in any given locality one is free to transform to a frame of reference
with a vanishing local ``gravitational field'' (connection coefficients),
then any local definition of (changes in) the energy-momentum of that
field would likewise vanish (even in situations where such changes
are physically expected).

The problem of defining gravitational energy-momentum is a subtle
one, and its resolution still lacks a general consensus among relativists
today \cite{szabados_quasi-local_2004,jaramillo_mass_2011}. It is
nevertheless widely accepted that in the spatial infinity limit of
an asymptotically-flat vacuum spacetime, any proposals for such definitions
should recover the ADM definitions \cite{arnowitt_dynamics_1962,regge_role_1974}.
For example, the ADM energy $E_{\textrm{ADM}}$ of an asymptotically-flat
vacuum spacetime is given by the integral over a closed two-surface
$\mathscr{S}\simeq\mathbb{S}_{r}^{2}$ (topologically a two-sphere
$\mathbb{S}_{r}^{2}$ of areal radius $r$) at spatial infinity ($r\rightarrow\infty$)
of an \textit{energy surface density} (energy per unit area), given
up to a factor by the trace $k$ of the extrinsic curvature of that
surface, 
\begin{equation}
E_{\textrm{ADM}}=-\frac{1}{8\pi}\lim_{r\rightarrow\infty}\oint_{\mathscr{S}}\bm{\epsilon}_{\mathscr{S}}^{\,}\,k\,.\label{eq:E_ADM}
\end{equation}

The ADM definitions have proven widely useful in practice, but in
principle are limited to determining the gravitational energy-momentum
of an \textit{entire} (asymptotically-flat vacuum) spacetime. Various
current proposals exist \cite{szabados_quasi-local_2004,jaramillo_mass_2011}
for the gravitational energy-momentum of \textit{arbitrary} spacetime
regions within arbitrary spacetimes. These have generally retained
the basic mathematical form (with an exact recovery in the appropriate
limit) of the ADM definitions, i.e. that of closed two-surface integrals
(of \textit{surface densities}), and are for this reason referred
to as \textit{quasilocal} definitions.

\begin{figure}
\begin{centering}
\includegraphics[scale=0.9]{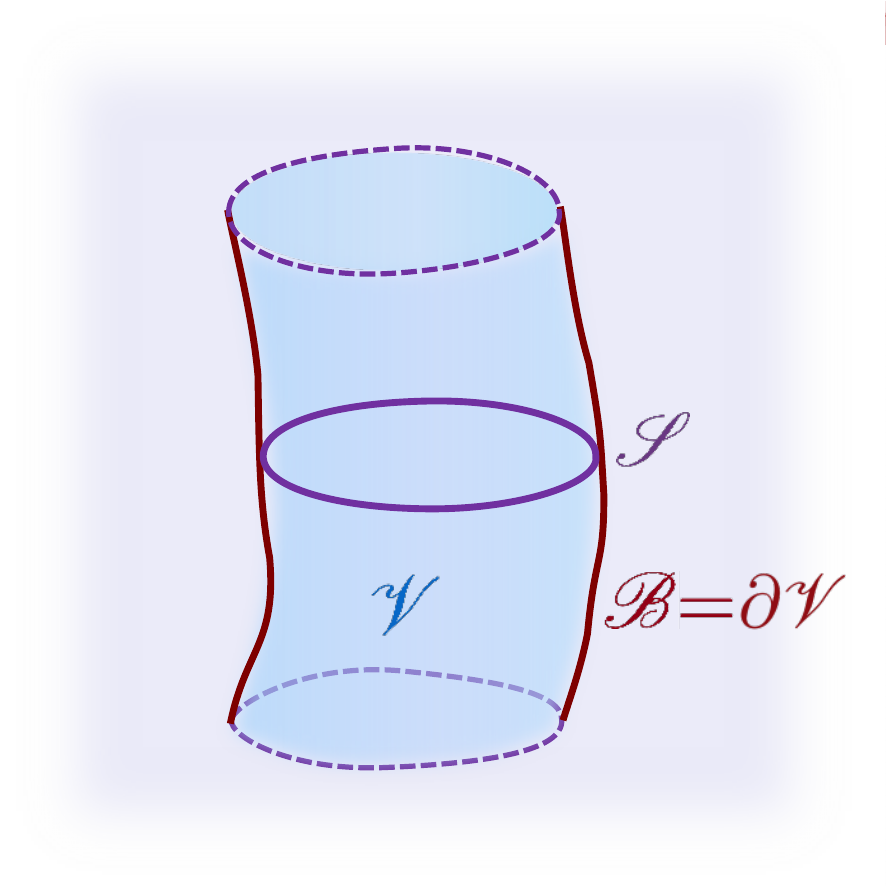} 
\par\end{centering}
\caption{A $(2+1)$ picture of a worldtube $\mathscr{V}$ with boundary $\mathscr{B}=\partial\mathscr{V}$.
A spatial slice of the latter is denoted by $\mathscr{S}$ (a closed
two-surface, topologically a two-sphere).}
\label{fig-qc-V} 
\end{figure}

In this essay, we assume and work with the quasilocal stress-energy-momentum
tensor proposed by Brown and York \cite{brown_quasilocal_1993}. Its
definition can simply be motivated by the following argument \cite{epp_momentum_2013}.
Recall that the stress-energy-momentum tensor of matter \textit{alone}
is defined from the matter action $S_{\textrm{matter}}$, up to a
factor, as $T_{ab}\propto\delta S_{\textrm{matter}}/\delta g^{ab}$,
which is a local tensor (living in the bulk). Following a similar
logic, consider a \textit{total} (matter plus gravitational) action,
\begin{equation}
S_{\textrm{total}}=S_{\textrm{matter}}+S_{\textrm{gravity}}\,.\label{eq:S_total}
\end{equation}
The gravitational action $S_{\textrm{gravity}}$ is, in any spacetime
region $\mathscr{V}$—which for simplicity henceforth we take to be
a worldtube\footnote{\setstretch{0.9}This assumption is made here only to simplify our
discussion. For a full analysis for a completely arbitrary $\mathscr{V}$,
see e.g. \cite{brown_action_2002}.}, i.e. the history of a finite spatial three-volume, see Fig. \ref{fig-qc-V}—as
a sum, 
\begin{equation}
S_{\textrm{gravity}}=S_{\textrm{EH}}+S_{\textrm{GHY}}\,.\label{eq:S_gravity}
\end{equation}
The first term is the Einstein-Hilbert (bulk) term, 
\begin{equation}
S_{\textrm{EH}}=\frac{1}{16\pi}\int_{\mathscr{V}}\bm{\epsilon}_{\mathscr{V}}^{\,}\,R\,,\label{eq:S_EH}
\end{equation}
and the second is the Gibbons-Hawking-York (boundary) term, 
\begin{equation}
S_{\textrm{GHY}}=-\frac{1}{8\pi}\int_{\mathscr{B}}\bm{\epsilon}_{\mathscr{B}}^{\,}\,K\,,\label{eq:S_GHY}
\end{equation}
where $K$ is the trace of the extrinsic curvature of the boundary
$\mathscr{B}=\partial\mathscr{V}\simeq\mathbb{R}\times\mathbb{S}^{2}$. From the total action (\ref{eq:S_total}),
a \textit{total} stress-energy-momentum tensor $\tau_{ab}$ can be
defined (analogously to the matter-only $T_{ab}$), as $\tau_{ab}\propto\delta S_{\textrm{total}}/\delta g^{ab}$.
Assuming the Einstein equation (\ref{eq:EE}) holds in $\mathscr{V}$,
the bulk term in the functional derivative now vanishes, and the result
evaluates to a tensor \textit{living on the boundary} $\mathscr{B}$,
i.e. a quasilocal tensor, known as the \textit{Brown-York tensor},
and given (with the appropriate proportionality factor restored) by 
\begin{equation}
\tau_{ab}=-\frac{1}{8\pi}\Pi_{ab}\,,\label{eq:tau}
\end{equation}
where $\Pi_{ab}$ is the canonical momentum (defined in the usual
way from the extrinsic curvature) of $\mathscr{B}$. This expresses
boundary densities of the total (matter plus gravitational) energy-momentum
which, when integrated 
over the two-surface intersection of a spacelike Cauchy slice and $\mathscr{B}$, yield the total values thereof contained in the part of the Cauchy slice (three-volume) within $\mathscr{B}$.

\begin{figure}
\begin{centering}
\includegraphics[scale=0.9]{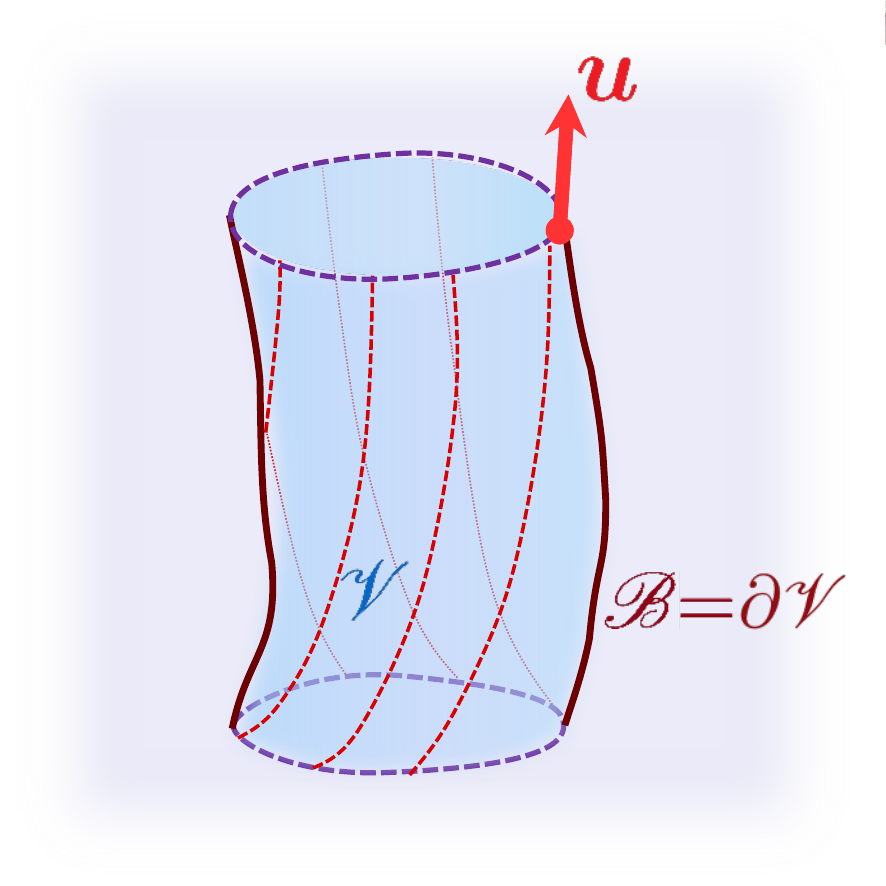} 
\par\end{centering}
\caption{A $(2+1)$ picture of a quasilocal frame $(\mathscr{B},u^{a})$, where
$\mathscr{B}=\partial\mathscr{V}$ is the boundary of a worldtube
$\mathscr{V}$ and $u^{a}$ is the timelike four-velocity of a two-parameter
family of observers the integral curves of which constitute $\mathscr{B}$.}
\label{fig-qc-Vqf} 
\end{figure}

Conservation laws for energy, momentum and angular momentum using
the Brown-York tensor have been formulated with the use of a concept
called \textit{quasilocal frames} \cite{mcgrath_quasilocal_2012,epp_momentum_2013}.
Essentially, the idea is that additional structure is required on
$\mathscr{B}$ in order to specify 
the components of stress-energy-momentum seen be a particular set of observers on $\mathscr{B}$. In particular, what is required is a two-parameter
congruence with timelike observer four-velocity $u^{a}\in T\mathscr{B}$, the
integral curves of which constitute $\mathscr{B}$. Such a pair $(\mathscr{B},u^{a})$
is referred to as a quasilocal frame. See Fig. \ref{fig-qc-Vqf}.
We thus have, e.g., the following general expression for the quasilocal
energy surface density (and similar expressions for the momentum and
stress): 
\begin{equation}
\mathcal{E}=\tau_{ab}u^{a}u^{b}=-\frac{1}{8\pi}k\,,\label{eq:Egeneral}
\end{equation}
where $k$ is the observers' (two-dimensional) spatial trace of the extrinsic curvature of $\mathscr{B}$.
In the simplest case of $u^{a}$ being orthogonal to a spatial slice $\mathscr{S}$ of $\mathscr{B}$ (which is sufficient for our purposes here), the total energy contained in the spatial volume inside $\mathscr{S}$ is: 
\begin{equation}
\mathtt{E}=\int_{\mathscr{S}}\bm{\epsilon}_{\mathscr{S}}^{\,}\,\mathcal{E}\,.\label{eq:Etotgeneral}
\end{equation}

The advantage of this construction is that it permits the computation
of (changes in) these various quantities for any region $\mathscr{V}$
on the boundary of which such a congruence (quasilocal frame) can
be defined. For any small spatial region, that is, one contained inside
a topological two-sphere having an areal radius $r$ much smaller
than the spacetime curvature and scale of matter density variation, the quasilocal energy density (\ref{eq:Egeneral})
evaluates in general to \cite{epp_momentum_2013}: 
\begin{equation}
\mathcal{E}=\mathcal{E}_{\textrm{vac}}+\mathcal{O}\left(r\right)\,,\label{eq:Esmallr}
\end{equation}
where 
\begin{equation}
\mathcal{E}_{\textrm{vac}}=-\frac{1}{4\pi r}\label{eq:Evac}
\end{equation}
is known as the \textit{vacuum} quasilocal energy density. Matter
contributions to $\mathcal{E}$ begin possibly from $\mathcal{O}(r)$
and gravitational contributions possibly from $\mathcal{O}(r^{3})$.
This vacuum energy is a geometrical term, simply accountable from
the fact that the extrinsic curvature trace of a round two-sphere in flat space
is $k=2/r$, and it is often regarded in the literature as unphysical
\cite{szabados_quasi-local_2004,brown_action_2002}. Yet, analyses
and applications of the quasilocal conservation laws have shown how
this term is in fact needed for a proper accounting of gravitational
energy-momentum transfer \cite{epp_momentum_2013}. In particular,
it is intimately linked to and logically self-consistent with the
existence of a \textit{vacuum pressure} (similarly, the leading term
in an expansion in $r$ of the quasilocal pressure ${\rm P}$, defined
from the observers' spatial trace of $\tau_{ab}$), 
\begin{equation}
{\rm P}_{\textrm{vac}}=-\frac{1}{8\pi r}\,.\label{eq:Pvac}
\end{equation}
Physically, a negative vacuum pressure acts as a positive surface
tension on the boundary $\mathscr{S}$, resulting in a ``${\rm P}_{\textrm{vac}} {\rm d}A$'' work term, which exactly accounts for the change in negative
vacuum energy. These vacuum terms have been shown to be necessary
in applications, including recently in the gravitational self-force
problem \cite{oltean_motion_2019}, where they play a key role in
accounting for the perturbative correction to the motion of a point
particle due to gravitational back-reaction.

Thus far, to our knowledge, these quasilocal quantities and their conservation laws
have not been investigated in detail in the context of cosmology.
In what follows, we present a basic application of these to a flat
FLRW spacetime with a perfect-fluid matter $T_{ab}$. We begin by
writing the metric $g_{ab}$ in spherical coordinates $\{X^{a}\}=\{T,R,\Theta,\Phi\}$
as 
\begin{equation}
g_{ab}{\rm d}X^{a}{\rm d}X^{b}=-{\rm d}T^{2}+a^{2}(T)\left[{\rm d}R^{2}+R^{2}{\rm d}\Omega^{2}\right]\,,\label{eq:FLRW}
\end{equation}
where ${\rm d}\Omega$ is the line element on the unit two-sphere
with coordinates $\{\Theta,\Phi\}$. We transform this now into coordinates
$\{x^{a}\}=\{t,r,\theta,\phi\}$ adapted to a time-dependent spherically-symmetric
quasilocal frame $(\mathscr{B},u^{a})$, such that $\mathscr{B}=\{r=\textrm{const}.\}$,
by applying the transformation 
\begin{equation}
\begin{cases}
T & =t\,,\\
R & =rf(t)\,,\\
\Theta & =\theta\,,\\
\Phi & =\phi\,,
\end{cases}\label{eq:coords}
\end{equation}
where $f(t)$ specifies the boundary time-dependence. Thus $\mathscr{S}=\mathbb{S}_{r}^{2}$
at any $t=\textrm{const}$. In these coordinates, for any $a(t)$
and $f(t)$, the four-velocity of the quasilocal observers is
\begin{equation}
u^{a}=\frac{1}{\sqrt{1-a^{2}\dot{f}^{2}r^{2}}}\delta^{a}\,_{0}\,,\label{eq:ua}
\end{equation}
and the quasilocal energy density (\ref{eq:Egeneral}), evaluates
to 
\begin{equation}
\mathcal{E}=-\frac{1+a\dot{a}f\dot{f}r^{2}}{4\pi afr\sqrt{1-a^{2}\dot{f}^{2}r^{2}}}\,.\label{eq:Egeneral_cosmo}
\end{equation}

\begin{figure}
\begin{centering}
\includegraphics[scale=1.1]{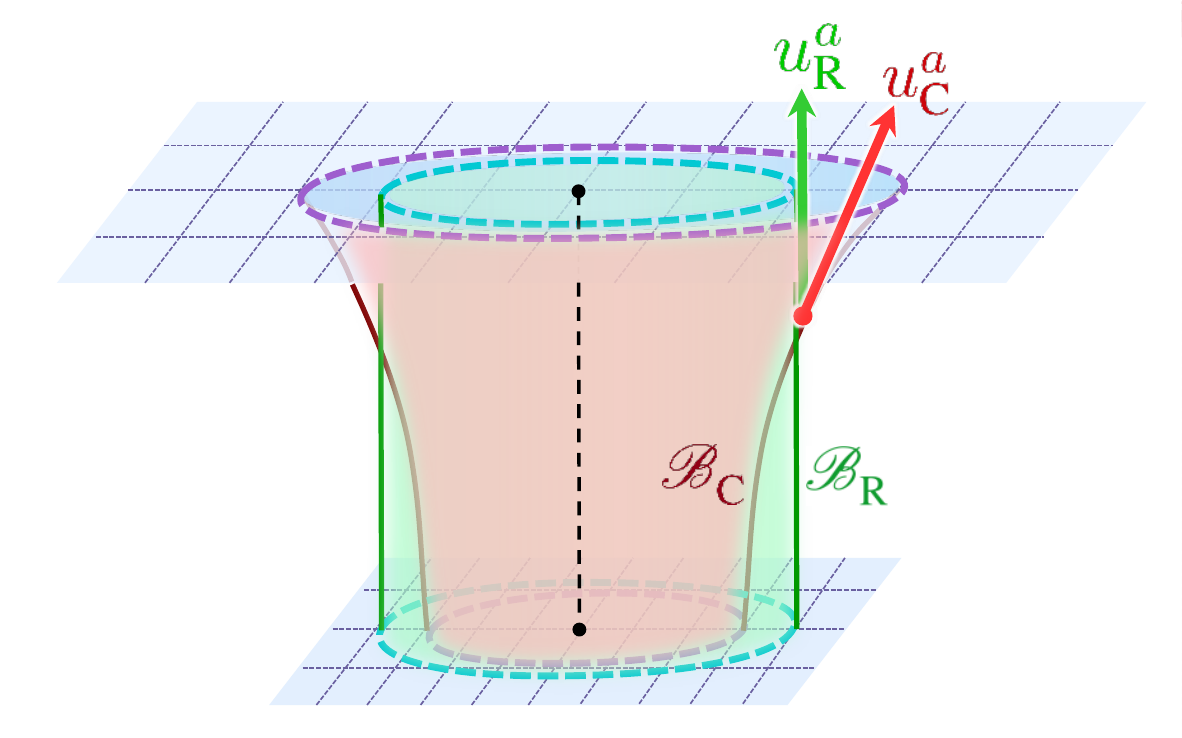} 
\par\end{centering}
\caption{A $(2+1)$ picture in a flat FLRW spacetime showing, in red, a co-moving
quasilocal frame $(\mathscr{B}_{\textrm{C}},u_{\textrm{C}}^{a})$
(with the boundary at a fixed co-moving radius) and, in green, a rigid
quasilocal frame $(\mathscr{B}_{\textrm{R}},u_{\textrm{R}}^{a})$
(with the boundary at a fixed proper radius). The black dotted line
is the center of the spatial FLRW coordinates $\{X^{a}\}$, depicted
(as a Cartesian system) on each Cauchy slice by dotted blue lines.}
\label{fig-qc-frw} 
\end{figure}

Two cases of interest are when the quasilocal observers are \textit{co-moving},
i.e. at a fixed coordinate radius (and so the spacetime physics is
essentially encoded in their motion), and when they are \textit{rigid},
i.e. at a fixed proper radius (and thus the physics is essentially
encoded in the energy-momentum boundary fluxes), respectively
given by the choices:
\begin{align}
f_{\textrm{C}}(t)\, & =1\,,\label{eq:fC}\\
f_{\textrm{R}}(t)\, & =1/a(t)\,.\label{eq:fR}
\end{align}
Let us denote these quasilocal frames
respectively as $(\mathscr{B}_{\textrm{C}},u_{\textrm{C}}^{a})$ and
$(\mathscr{B}_{\textrm{R}},u_{\textrm{R}}^{a})$. See Fig. \ref{fig-qc-frw}. The co-moving observers
see only the energy of the vacuum,
\begin{equation}
\mathcal{E}_{\textrm{C}}=-\frac{1}{4\pi ar}\,,\label{eq:EC}
\end{equation}
while for rigid observers, the quasilocal energy is
\begin{equation}
\mathcal{E}_{\textrm{R}}=-\frac{\sqrt{1-H^{2}r^{2}}}{4\pi r}\,,\label{eq:ER}
\end{equation}
where $H=\dot{a}/a$.

We can now calculate from these the total energies, given by (\ref{eq:Etotgeneral}).
For the co-moving energy, we find the conservation expression $\dot{\mathtt{E}}_{\textrm{C}}=H\mathtt{E}_{\textrm{C}}$,
telling us that this energy changes at an exponential rate of the integral of $H(t)$.
Moreover, the total rigid and co-moving energies are simply related
by a time dilation factor, $\mathtt{E}_{\textrm{R}}=\mathtt{E}_{\textrm{C}}\sqrt{1-v^{2}}$,
where $v=-rH$ is the relative radial velocity of rigid quasilocal observers
as seen by co-moving observers.

It is interesting to consider $\mathtt{E}_{\textrm{R}}$ for small
$r$ (in a ``small locality''). Denoting
the total (flat-space) vacuum energy as $\mathtt{E}_{\textrm{vac}}=\int_{\mathscr{S}}\bm{\epsilon}_{\mathscr{S}}^{\,}\,\mathcal{E}_{\textrm{vac}}=-r$, this evaluates to $\mathtt{E}_{\textrm{R}}=\mathtt{E}_{\textrm{vac}}+\frac{1}{2}H^{2}r^{3}+\mathcal{O}(r^{5})$.
Inserting the Friedmann equation $H^{2}=\frac{8\pi}{3}\rho+\frac{1}{3}\Lambda$,
this can be written as:
\begin{equation}
\mathtt{E}_{\textrm{R}}=\mathtt{E}_{\textrm{vac}}+\left(\rho+\rho_{\Lambda}\right)\frac{4\pi r^{3}}{3}+\mathcal{O}(r^{5})\,.\label{eq:EPexpansion}
\end{equation}
Thus at leading order beyond the vacuum, we recover the usual mass
of matter (as the three-ball volume multiplying the matter energy density
$\rho$), along with the cosmological constant term which can thus
be seen to behave, locally, similarly to matter with a (volume) energy density
$\rho_{\Lambda}=\frac{\Lambda}{8\pi}$. Concordantly, the quasilocal
pressure after using both Friedmann equations has the expansion
\begin{equation}
{\rm P}_{\textrm{R}}={\rm P}_{\textrm{vac}}-\frac{1}{2}\left(p+p_{\Lambda}\right)r+\mathcal{O}(r^{3})\,,\label{eq:PRexpansion}
\end{equation}
where $p_{\Lambda}=-\frac{\Lambda}{8\pi}=-\rho_{\Lambda}$, so the
cosmological constant term also acts locally like matter with a pressure
opposite in sign to its effective volume energy density.

These results offer a basic initial confirmation that these quasilocal
methods yield reasonable results when applied to cosmology, and good
encouragement to develop this application further. In particular,
the perspective that may be gained upon the nature of the quasilocal
vacuum energy and its relation to the cosmological constant warrants
further conceptual investigation. While we have presented a simple
computation in this essay to illustrate the proof of concept, we are
working on also including and deriving conservation laws for cosmological
perturbations using these methods \cite{quasicosmo-forthcoming}.

~

\noindent\textit{Acknowledgements.} We thank Robert H. Brandenberger for useful discussions and comments on an early draft of this essay.


%

\end{document}